\documentclass[aps,prd,twocolumn,superscriptaddress]{revtex4-1}

\pdfoutput=1
\usepackage{ifpdf}
\usepackage[pdftex]{graphicx} 
\usepackage{color} 

\usepackage{amsfonts}
\usepackage{amsmath}
\usepackage{bm}
\usepackage{mathrsfs}
\usepackage{here}
\newif\ifoneauthor
\oneauthortrue
\newcommand{\unit}[1]{\ {\rm #1}}

\newcommand{\Section}[1]{section \ref{#1}}
\newcommand{\Eq}[1]{Eq. (\ref{#1})}
\newcommand{\Table}[1]{Table \ref{#1}}
\newcommand{\Fig}[1]{Figure \ref{#1}}
\DeclareMathAlphabet{\mathpzc}{OT1}{pzc}{m}{it}
\usepackage{here} 
\definecolor{gray}{gray}{0.4}

\begin{document}

\title{Prospects for gravitational-wave polarization test \\ from compact binary mergers with future ground-based detectors}


\author{Hiroki Takeda}
\email[]{takeda@granite.phys.s.u-tokyo.ac.jp}
\affiliation{Department of Physics, University of Tokyo, Bunkyo, Tokyo 113-0033, Japan}
\author{Atsushi Nishizawa}
\affiliation{Research Center for the Early Universe (RESCEU), School of Science, University of Tokyo, Bunkyo, Tokyo 113-0033, Japan}
\affiliation{Kobayashi-Maskawa Institute for the Origin of Particles and the Universe,Nagoya University, Nagoya, Aichi 464-8602, Japan}
\author{Koji Nagano}
\affiliation{KAGRA Observatory, Institute for Cosmic Ray Research, University of Tokyo, Kashiwa, Chiba, 277-8582, Japan}
\author{Yuta~Michimura}
\affiliation{Department of Physics, University of Tokyo, Bunkyo, Tokyo 113-0033, Japan}
\author{Kentaro Komori}
\affiliation{Department of Physics, University of Tokyo, Bunkyo, Tokyo 113-0033, Japan}
\author{Masaki Ando}
\affiliation{Department of Physics, University of Tokyo, Bunkyo, Tokyo 113-0033, Japan}
\author{Kazuhiro Hayama}
\affiliation{Department of Applied Physics, Fukuoka University, Nanakuma, Fukuoka 814-0180, Japan}




\date{\today}

\begin{abstract}
There exist six possible polarization modes of gravitational waves in general metric theory of gravity, while two tensor polarization modes are allowed in general relativity. The properties and number of polarization modes depend on gravity theories. The number of the detectors needs to be equal to the number of the polarization modes of the gravitational waves for separation of polarizations basically. However, a single detector having great sensitivity at lower frequency could be effectively regarded as a virtual detector network including a set of detectors along its trajectory due to a long gravitational-wave signal from a compact binary and the Earth's rotation. Thus, time-varying antenna pattern functions can help testing the polarizations of gravitational waves. We study the effects of the Earth's rotation on the polarization test and show a possibility to test the nontensorial polarization modes from future observations of compact binary mergers with ground-based gravitational detectors such as Einstein telescope and Cosmic Explorer.
\end{abstract}

\pacs{42.79.Bh, 95.55.Ym, 04.80.Nn, 05.40.Ca}

\maketitle

\section{Introduction}
 The observations of gravitational waves (GWs) from compact binary coalescences (CBC) \cite{Abbott2016c, TheLIGOScientificCollaboration2018b} by Advanced LIGO (aLIGO) \cite{Aasi2015} and Advanced Virgo (AdV) \cite{Acernese2015} succeeded in acquiring experimental access to the nature of gravity and spacetime. Several investigations to test general relativity (GR) by the observations of GWs have been proposed and carried out \cite{Abbott2016g, Abbott2016e, Abbott2017, Abbott2017b, Callister2017}, for example, tests of post-Newtonian gravity,  analysis of quasi-normal modes, and probing into the dispersion relation of GW. No evidence for violations of GR have been reported before. It is expected that the GW detector network will expand soon by participation of other ground-based detectors such as KAGRA  \cite{Akutsu2018, KAGRACollaboration2019} and LIGO India \cite{Unnikrishnan2015}. Thus, more precise tests of gravity would be possible in the near future.

It is well known that a GW has only two tensor polarization modes (plus, cross) in general relativity by fixing gauge degrees of freedom between a perturbed spacetime and a background spacetime \cite{Misner1973, Will2005, Maggiore2007, Creighton2011}. However, a generic metric theory allows at most six polarizations: two tensor modes (plus, cross), two vector modes (vector x, vector y), and two scalar modes(breathing, longitudinal) \cite{Eardley1973a, Will1993}. The properties of the polarization modes of a GW reflect degrees of freedom or the gauge conditions in the theory of gravity.  Additional degrees of freedom of the theory or the breaking of the gauge symmetries result in leading to additional degrees of freedom for a GW. The possible polarization modes are reported  in each theory of gravity . For example, GWs in modified gravity theories such as scalar-tensor theory \cite{Brans1961, Fujii2003} and f(R) gravity \cite{Buchdahl1970, DeFelice2010, Sotiriou2010} can have scalar polarizations in addition to tensor modes \cite{Eardley1973a, Will1993, Will2005,  Hou2018}. Polarizations of null GWs can be treated by the Newman-Penrose formalism transparently \cite{Newman1962a, Alves2010a}. In contrast, up to six polarizations are possible \cite{Alves2010a} in bimetric gravity theory \cite{Visser1997, Hassan2011} while up to five polarizations are possible \cite{DeRham2011} in massive gravity theory \cite{Rubakov2008, DeRham2010}. A discovery of nontensorial polarization modes indicates the existence of an alternative to GR as a fundamental theory of gravity. Thus, a search for the polarizations of a GW is crucial and useful matter to approach to the essence of gravitation.

Some analytical attempts and studies to separate the polarization modes were made for GW bursts \cite{Hayama2013a}, stochastic GWs \cite{Nishizawa2009a}, continuous GWs \cite{Isi2015}, and GWs from compact binary coalescences \cite{Takeda2018}. In the previous work \cite{Takeda2018} for GWs from compact binary coalescences, we found the condition to separate the polarization modes that {\it the number of the detectors needs to be equal to the number of the polarization modes of the GW in principle.}

Currently, there are few observational constraints about the nontensorial polarization modes of GWs.  The radiated energy in scalar polarization modes have been limited to less than $\sim1\%$ by the observation of PSR B1913+16 \cite{Will2005}. A polarization mode search of GW170814 involved with the simple substitution of the antenna pattern functions has already been conducted \cite{Abbott2017b}. 
 
\begin{figure}
\begin{center}
\includegraphics[width=\hsize]{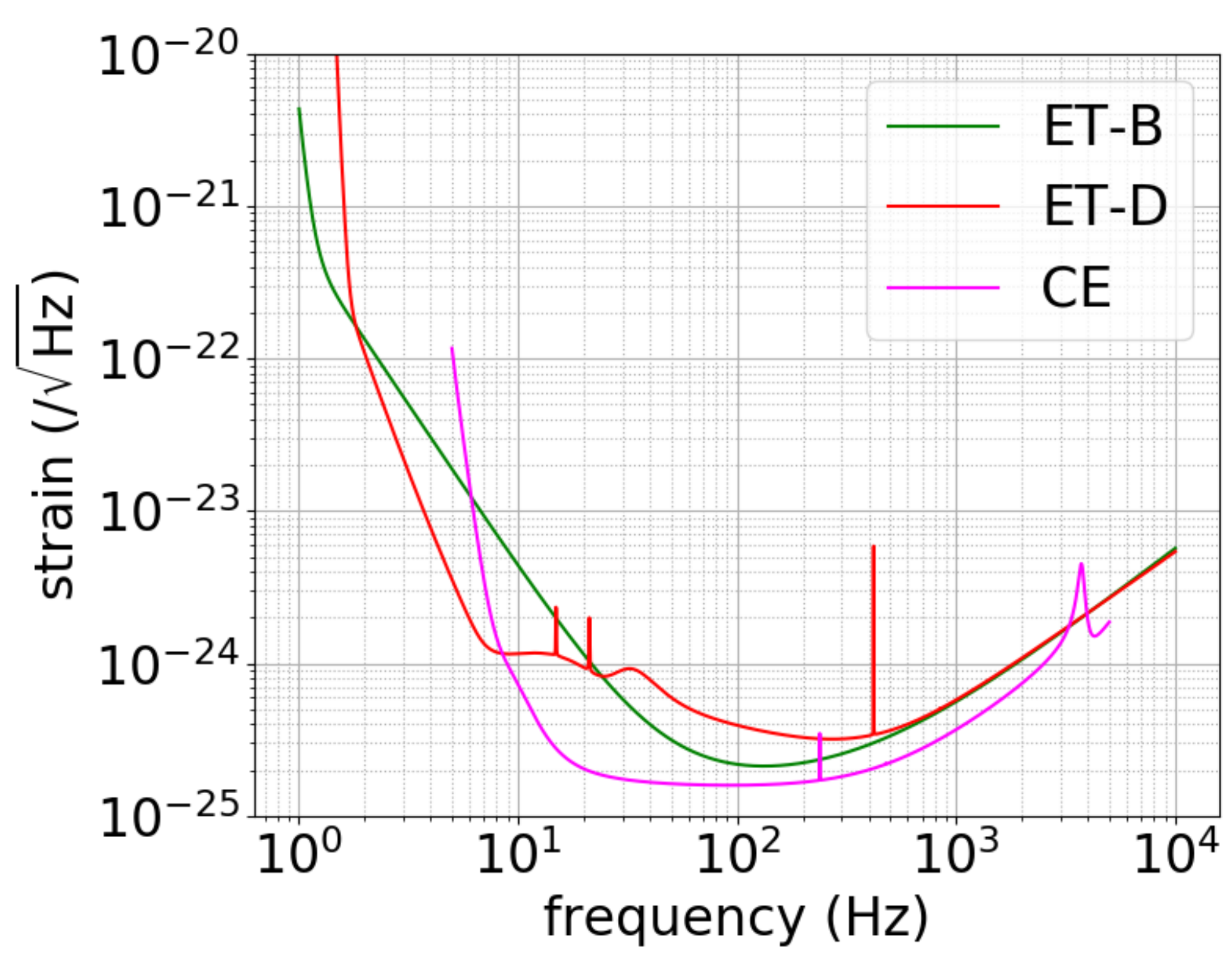}
\end{center}
\caption{Sensitivity curves for the third-generation gravitational-wave detectors such as Einstein telescope and Cosmic Explorer. Einstein telescope has two sensitivity estimates, ET-B and ET-D.}
\label{sensitivity_3G}
\end{figure}

We focus on the polarization test of a GW from CBC with future ground-based detectors. There are two leading designs of the next-generation GW detectors followed by so-called the second generation (2G) GW detectors such as aLIGO, AdV and KAGRA. One is the Einstein Telescope (ET) \cite{Punturo2010a} and the other is the Cosmic Explorer (CE) \cite{Abbott2016h}. These  detectors are called the third-generation (3G) GW detectors. \Fig{sensitivity_3G} shows their sensitivity curves. ET will have three 10km-arms in an equilateral triangle, with multiple interferometers sharing the arms. ET has two sensitivity estimates, ET-B and ET-D. ET-B is based on a single interferometric detector covering the frequency range from $1 {\rm Hz}$ to $10 {\rm kHz}$ \cite{Hild2008, Punturo2010}.  ET-D has the so-called xylophone design where it consists of one cryogenic low-frequency interferometer and one room-temperature high-frequency interferometer at each corner \cite{Hild2011}. The sensitivity around sub-10 Hz of ET-D is improved well in comparison with ET-B. CE has an L-shaped configuration like the 2G detectors. However, its arm length is $40\unit{km}$. Then, the sensitivity of CE is significantly better above 10 Hz than those of ET detectors.

Compact binary coalescences are promising sources for both 2G and 3G detectors. It is expected that 3G detectors can detect a lot of compact binary coalescences and determine their parameters more accurately because of their great sensitivity. We research prospects for the GW polarization test from compact binary mergers with the future ground-based detectors, i.e. 3G detectors. The number of the detectors needs to be more than or equal to the number of the polarization modes of the GW for the polarization test with the 2G detectors. However, several 3G detectors or even a single 3G detector may be able to test more polarizations than detectors, because 3G detectors such as ET and CE have substantially better sensitivity in low frequency region around $5\unit{Hz}$ than 2G detectors. Therefore, the rotation of the Earth and the detector positions changed with time need to be considered properly in the case of 3G detectors. It has been reported that the Earth rotational effect of an individual 3G detector helps localize the compact binary coalescences effectively due to time-dependent antenna pattern functions \cite{Zhao2017, Chan2018}. The effect of the Earth rotation is expected to be useful in the polarization test because time-varying antenna pattern functions lead to an effective detector network formed by a single detector moving in time due to the Earth's rotation. We study the polarization separability for binary black hole (BBH) and binary neutron star (BNS) with 3G detectors  through the determination accuracy of polarization parameters. Then, we assume a successive observation by 3G detectors and study prospects for the GW polarization test with multiple compact binary sources by considering a binary merger distribution.
This paper is organized as follows. In \Section{Detector_signal}, we describe polarization modes of gravitational waves, time-dependent antenna pattern functions and detector signals. In \Section{Polarization_models}, we explain models for a detector signal including nontensorial polarization modes of a GW. In \Section{Setup},  we introduce the basics of the Fisher analysis and our analytical and numerical setup. In \Section{Results}, we show our results of parameter estimation with 3G detectors and mention about separability of the polarizations and the Earth rotational effect in the polarization test for BNS and BBH. Then, we show prospects for the polarization test of GWs from multiple compact binary sources with future ground-based detectors. We devote the last \Section{Conclusion} to the conclusion of this paper. Throughout this paper we use units in which $G = c = 1$.

\section{Detector signal}
\label{Detector_signal}
\subsection{Polarization mode of  gravitational waves}
The most general null gravitational wave has six independent polarization modes\cite{Eardley1973a, Will1993}. Thus, a metric perturbation representing a GW can be expressed as
\begin{equation}
\label{gw}
h_{ab}(t,\hat{\Omega})=h_{A}(t)e^{A}_{ab}(\hat{\Omega}),
\end{equation}
at a point on a space-time manifold. Here polarization indices $A=+,\times, x, y, b, l$ denote plus, cross, vector x, vector y, breathing, and longitudinal polarization modes, respectively. $h_{A}(t)$ are the components of the GW for each polarizations. $\hat{\Omega}$ is the direction to the position of a GW source in the sky and polarization tensors  $e^{A}_{ab}(\hat{\Omega})$ are defined as bellow,
\begin{equation}
e^{+}_{ab}=\hat{e}_{x}\otimes\hat{e}_{x}-\hat{e}_{y}\otimes\hat{e}_{y},
\end{equation}
\begin{equation}
e^{\times}_{ab}=\hat{e}_{x}\otimes\hat{e}_{y}+\hat{e}_{y}\otimes\hat{e}_{x},
\end{equation}
\begin{equation}
e^{x}_{ab}=\hat{e}_{x}\otimes\hat{e}_{z}+\hat{e}_{z}\otimes\hat{e}_{x},
\end{equation}
\begin{equation}
e^{y}_{ab}=\hat{e}_{y}\otimes\hat{e}_{z}+\hat{e}_{z}\otimes\hat{e}_{y},
\end{equation}
\begin{equation}
e^{b}_{ab}=\hat{e}_{x}\otimes\hat{e}_{x}+\hat{e}_{y}\otimes\hat{e}_{y},
\end{equation}
\begin{equation}
e^{l}_{ab}=\sqrt{2}\hat{e}_{z}\otimes\hat{e}_{z}.
\end{equation}
The set of three unit vectors $\{\hat{e}_x, \hat{e}_y, \hat{e}_z\}$ forms the wave orthonormal coordinate such that $\hat{e}_z=\hat{e}_x\times \hat{e}_y$ and $\hat{e}_z=-\hat{\Omega}$ becomes a unit vector  pointing to the propagation direction of the GW. A degree of freedom to choose  $\hat{e}_x, \hat{e}_y$ around the $\hat{e}_z$ axis remains. This degree of freedom is referred as the polarization angle $\psi_p$. 

\subsection{Time-dependent antenna pattern functions and detector signal}
 The detector signal of the I-th GW detector are given by \cite{Tobar1999, Nishizawa2009a, Hayama2013a}
 \begin{equation}
 \label{detector_signal}
 h_I(t,\hat{\Omega})=d_{I}^{ab}(t)h_{ab}(t,\hat{\Omega})=F_I^{A}(t,\hat{\Omega})h_A(t),
 \end{equation} 
 where $d_I^{ab}$ is the detector tensor defined for each GW detector. 
 The detector tensor of an interferometric detector can be defined as
  \begin{equation}
d_I^{ab}(t):=\frac{1}{2}(\hat{u}_{I}^{a}(t)\otimes\hat{u}_{I}^{b}(t)-\hat{v}_{I}^{a}(t)\otimes\hat{v}_{I}^{b}(t)).
  \end{equation}
Here $\hat{u}_I, \hat{v}_I$ are unit vectors along the arms of the I-th interferometric detector.
  
$ F_{I}^A$ is the antenna pattern functions of the I-th detector for polarization "A" defined by
  \begin{equation}
  \label{antenna}
  F_I^{A}(t,\hat{\Omega}):=d_I^{ab}(t)e^{A}_{ab}(\hat{\Omega}).
  \end{equation}
  They represent the detector responses to each polarization mode. In general, two unit vectors $\hat{u}_I(t), \hat{v}_I(t)$ depend on time because the position of the detector changes in time due to the Earth's rotation or revolution. It results in the time-varying detector tensor and antenna pattern functions. We can regard the antenna pattern functions of the 2G GW detectors such as aLIGO, AdV and KAGRA as constants in time because a GW signal in the observational band is short so that the time dependence can be ignored. However, since the 3G GW detectors such as ET and CE or the space GW detectors such as LISA \cite{Amaro-Seoane2017, Armano2018}, DECIGO \cite{Kawamura2011} and Tianqin \cite{Luo2016} have extended lower frequency sensitivities, the time dependence can not be ignored. The general concrete formulas of the antenna pattern are provided in \cite{Nishizawa2009a} when the detector tensor does not depend on time. 
  
  \subsection{Fourier components of the detector signal}
In \Eq{detector_signal}, we shall consider the two tensor polarization modes, i.e. polarization index runs over $\{+, \times\}$ and the inspiral waveform as $h_A(t)$. The detector signal of inspiral GWs \Eq{detector_signal} from compact binary coalescences in time domain can be expressed as follows \cite{Creighton2011, Berti2005},
\begin{equation}
\label{signal_time}
h(t)\simeq\frac{2m_{1}m_{2}}{r_s(t)D_L}\mathscr{A}(t)\cos{(\int^tf_{\rm gw}(t')dt'+\phi_p(t)+\phi_D(t))},
\end{equation}
where $\mathscr{A}(t)$ and $\phi_p(t)$ are defined as below,
\begin{equation}
\mathscr{A}(t):=\sqrt{(1+\cos^2{\iota})^2F^{+}(t)^2+4\cos^2{\iota}F^{\times}(t)^2},
\end{equation}
\begin{equation}
\phi_p(t):=\arctan\left(\frac{2\cos{\iota}F^{\times}(t)}{(1+\cos^2{\iota})F^{+}(t)}\right).
\end{equation}
Here $m_1, m_2$ are the masses of compact binary stars, $D_L$ is the luminosity distance to the binary system, $r_s(t)$ is the orbital relative distance, $f_{\rm gw}$ is the frequency of the GW and $\phi_D(t)$  is the doppler phase.

We derive the Fourier components $h(f)$ of the detector signal $h(t)$. The Fourier component of the measured signal can be evaluated by the stationary phase approximation, since $(2m_{1}m_{2})/({r_s(t)D_L}), \mathscr{A}(t), \phi_p, \phi_D$ vary in time slowly. Employing stationary phase approximation, we can estimate the Fourier components of the detector signal \cite{Cutler1997, Berti2005, Arun2006, Maggiore2007, Zhao2017},
 \begin{equation}
 h_I(f)=\mathcal{A}f^{-7/6}e^{i\Psi(f)}\left\{\frac{5}{4}\mathscr{A}(t(f))\right\}e^{-i(\phi_p(t(f))+\phi_D(t(f)))}.
 \end{equation}
 The geometrical factor for tensor modes is defined by
  \begin{eqnarray}
 \label{geo_tensor}
 \mathcal{G}_{T,I}:=\frac{5}{2}\{(1+\cos^2{\iota})F_{+,I}(t)\nonumber \\ 	+2i\cos{\iota}F_{\times,I}(t)\}
 e^{i \phi_{D,I}(\theta_s,\phi_s,\theta_e,\phi_e)},
 \end{eqnarray}
 where $(\theta_s,\phi_s)$ are the the source direction angular parameters and $(\theta_e,\phi_e)$ are detector position parameters.
The factor of $5/2$ appears such that the average of \Eq{geo_tensor} over angular parameters gives unity.
Then we get \begin{equation}
\label{signal_tensor}
 h_I(f)=\mathcal{A}f^{-7/6}e^{i\Psi(f)}\mathcal{G}_{T,I}(t(f)).
\end{equation}


\noindent
 $t(f)$ is 
 \begin{equation}
 \label{tf}
 t(f):=t_c-\frac{5}{256}\mathcal{M}^{-5/3}(\pi f)^{-8/3},
 \end{equation} 
 such that  the condition,
\begin{equation}
f=f_{\rm gw}(t(f)),
\end{equation}
is satisfied. Here $\mathcal{M}$ is the chirp mass and $t_c$  is the coalescence time. $t(f)$ gives the relation between the time and the frequency of the GW before merger.

\section{Polarization models}
\label{Polarization_models}
We mention the polarization models used in our analysis. We assume that nontensorial polarization modes have the same waveforms as the tensor mode $h_{\rm GR}$, though these waveforms actually depend on a specific theory of gravity. This means we consider pessimistic cases in terms  of separation of the polarization modes as far as we do not introduce any specific parameters of the theory because it is more difficult to separate the modes having the same waveforms. The shortness of the signal could affect the separability, especially in the case of BBH \cite{Takeda2018}. However, since a radiation process in merging and ringdown phase in a modified gravity is complicated, it is difficult in general to relate the polarizations in inspiral phase to those in merger and ringdown phases. Here we focus only on an inspiral phase to be conservative and to keep results robust.

The signal models for scalar and vecotr polarization modes have been derived in Eq.(29)-Eq.(34) of \cite{Takeda2018}. However, the time dependence of detectors was not taken into account. To extend them and obtain time-dependent polarization models, we can substitute \Eq{tf} into the geometrical factors as we derived \Eq{signal_tensor}. We consider only the following polarization model (Model TS1) having an additional scalar polarization mode in addition to two tensor polarization modes because  we already showed in our previous work \cite{Takeda2018} that a choice of polarizations causes only the different angular dependences in the signals and qualitatively small differences in the results. 

Model TS1 is a tensor-scalar dipole model. In this model, we add a scalar mode having the same inclination-angle dependence as that of dipole radiation.
An additional model parameter is the amplitude of the scalar mode $A_{S_1}$. 
\begin{equation}
h_I=\{\mathcal{G}_{T,I}(t(f))+A_{S_1}\mathcal{G}_{S_1,I}(t(f))\}h_{\rm{GR}},
\end{equation}
where $\mathcal{G}_{S_1,I}$ is the geometrical factor for the scalar mode for $I$-th detector, defined by
\begin{equation}
    \mathcal{G}_{S_1,I}:=\sqrt{\frac{45}{2}}\sin{\iota}F_{b,I}(\bm{\theta_s},\bm{\theta_e})e^{i \phi_{D,I}(\theta_s,\phi_s,\theta_e,\phi_e)}.
\end{equation}  
Here $\bm{\theta_s}:=(\theta_s,\phi_s,\psi_p)$ is a set of source angle parameters where $\psi_p$ is polarization angle and $\bm{\theta_e}:=(\theta_e,\phi_e, \psi)$ is  a set of detector angle parameters where $\psi$ specifies the detector orientation. Geometrical factors are normalized by angular average over the whole-sky and the inclination angle of the binary system.

\section{Setup}
\label{Setup}
\subsection{Fisher Analysis} 
  We evaluated the model parameter estimation by a Fisher information matrix \cite{Finn1992, Cutler1994,Creighton2011}. A Fisher information matrix tells us how precisely we can determine the model parameters by observations and how strongly the model parameters are correlated.
The Fisher information matrix $\Gamma$ can be calculated by
\begin{equation}
\Gamma_{ij}:=4{\rm{Re}}\int^{\rm{f_{max}}}_{\rm{f_{min}}}df\sum_I \frac{1}{S_{n,I}(f)}\frac{\partial h^{*}_I(f)}{\partial\lambda^i}\frac{\partial h_I(f)}{\partial\lambda^j},
\label{Fisher}
\end{equation} 
 where $S_{n,I}(f)$ is the I-th detector noise power spectrum and $\lambda_i$ is the i-th model parameter. 
The root mean square of a parameter can be given by the inverse of the Fisher information matrix. The root mean square of $\Delta\lambda^i$ is calculated as follows,
 \begin{equation}
(\Delta\lambda_i)_{\rm rms}:=\sqrt{\langle\Delta\lambda^i\Delta\lambda^i\rangle}=\sqrt{(\Gamma^{-1})^{ii}},
 \end{equation}
 where $\Delta\lambda^i$ is the measurement error of $\lambda^i$ and $\langle\cdot\rangle$ stands for ensemble average.
 
 The sky localization error of the source is defined by
 \begin{equation}
 \Delta\Omega_s:=2\pi|\sin{\theta_s}|\sqrt{\langle(\Delta\theta_s)^2\rangle\langle(\Delta\phi_s)^2\rangle-\langle\Delta\theta_s\Delta\phi_s\rangle^2}.
 \end{equation}

We simply refer to $(\Delta\lambda_i)_{\rm rms}$ as $\Delta\lambda_i$, and call it the estimation error of $\lambda_i$.

\subsection{Analytical and Numerical setups}
We adopt the inspiral waveform up to Newtonian order in amplitude and 3.5  post-Newtonian (PN) order  in phase \cite{Khan2016}
 \begin{equation}
 h_{\rm{GR}}=\mathcal{A}f^{-7/6} e^{i\Psi(f)},
 \end{equation}
 with
\begin{equation}
\mathcal{A}f^{-7/6}=\frac{1}{\sqrt{6}\pi^{2/3}d_L}\mathcal{M}^{5/6}f^{-7/6},
\end{equation}
\begin{equation}
\Psi(f)=2\pi ft_c-\phi_c-\frac{\pi}{4}+\frac{3}{128}(\pi\mathcal{M}f)^{-5/3}\sum_{i=0}^{7}\phi_i(\pi\mathcal{M}f)^{i/3}.
\end{equation}
The phase part of the above waveform includes the higher PN effect up to 3.5 PN order, while the amplitude part of the waveform is kept up to the Newtonian order because we consistently use the waveform at the same order as the expression in \Eq{tf}. 
The lower frequency end of the integration in \Eq{Fisher} is set to be $f_{\rm{min}}=1 \unit{Hz}$ and the upper frequency end $f_{\rm{max}}$  to be  the frequency $f_{\rm{ISCO}}$ that is twice the innermost stable circular orbit frequency for a point mass in Schwarzschild spacetime
\begin{equation}
f_{\rm{ISCO}}=(6^{3/2}\pi M_{\rm tot})^{-1},
\end{equation}
where $M_{\rm{tot}}=m_1+m_2$ is the total mass of the binary stars.

 We have 11 model parameters in GR
 \begin{equation}
(\log\mathcal{M},\log{\eta}, t_c, \phi_c, \log{d_L}, \chi_s, \chi_a, \theta_s, \phi_s, \cos{\iota}, \psi_p),
\end{equation}
 and an additional  polarization amplitude parameter $A_{S_1}$ in the model TS1.  Here, $\log{\eta}, \chi_s, \chi_a$ are the logarithm of the mass ratio, the symmetric spin parameter and the antisymmetric parameter, respectively. We assume that the fiducial values of $t_c, \phi_c, \chi_s, \chi_a$ are 0. This corresponds that the coalescence time and the phase at the coalescence time for the nontensorial mode are assumed to be those of the tensor mode because we showed that  the modification of the coalescence time and the phase at coalescence do not affect the results in our previous work. It is assumed that the fiducial values of the additional amplitude parameters are unity unless otherwise noted. The priors are imposed for parameters having domain of definition; $\log{\eta}$, $\phi_c$, angular parameters $(\theta_s, \phi_s, \cos{\iota}, \psi_p)$ and binary spin parameters $(\chi_s, \chi_a)$.
 
3G detectors are assumed to have their design sensitivity in \Fig{sensitivity_3G} and be positioned at Livingston site of aLIGO or Virgo site in the same way as \cite{Vitale2016}.
 
 We say that the polarization modes would be separable when the errors of the amplitudes parameter are less than their fiducial values. 
 
\section{Results}
\label{Results}
Here we show the possibility of the polarization test of GWs from compact binary mergers with 3G GW detectors and investigate how the detector sensitivity at low frequency region help with the polarization test.

\subsection{Binary neutron stars}
\label{BNS}
We study the ability of the 3G GW detectors in polarization test of GWs from compact binary coalescences.

First we estimate model parameters for BNS with equal $1.4 M_{\odot}$ at z=0.1 in the model TS1 with a single 3G GW detector such as ET-B, ET-D, CE and an ideal detector. The ideal detector has a better low-frequency constant sensitivity (with power spectral density $10^{-49}\ {\rm 1/Hz}$) than ET-D below $400\ {\rm Hz}$ to check the benefit of better low-frequency sensitivity in the polarization test of GWs. Angular parameters ($\cos{\theta_s},\phi_s,\cos{\iota},\psi_p$) are uniformly random and 500 sources are calculated by the Fisher analysis.

\begin{table*}
\caption{Medians of parameter estimation errors for $1.4M_{\odot}-1.4M_{\odot}$ BNS at $z=0.1$ with a single 3G GW detector. We say that the polarization modes would be separable when the errors of the amplitudes parameter are less than the fiducial value (Here, $A_{S1}=1$). Three polarization modes would be separable even with either an individual ET-D and an ideal detector.}
  \begin{tabular}{|c|c|c|c|c|c|} \hline
   & parameter & BNS(ET-B) & BNS(ET-D) & BNS(CE) & BNS(Ideal) \\ \hline
		 &SNR 							& 57.8 	& 50.7 		& 104 	& 170	\\  \hline 		  
		&$\Delta\ln{d_L}$ 					& 0.979 	&  0.355 		& 6.67 	& 0.197	\\ 
ModelTS1	&$\Delta\Omega_s[\rm{deg}^2]$ 		& 490 	&  55.6  		& 72105 	& 7.56	\\    
		&$\Delta A_{S1}$					& 1.30 	& 0.459 		& 12.9 	& 0.322	\\  \hline
  \end{tabular}
  \label{table_result}
\end{table*}

\begin{figure}
 \begin{center}
 \includegraphics[width=\hsize]{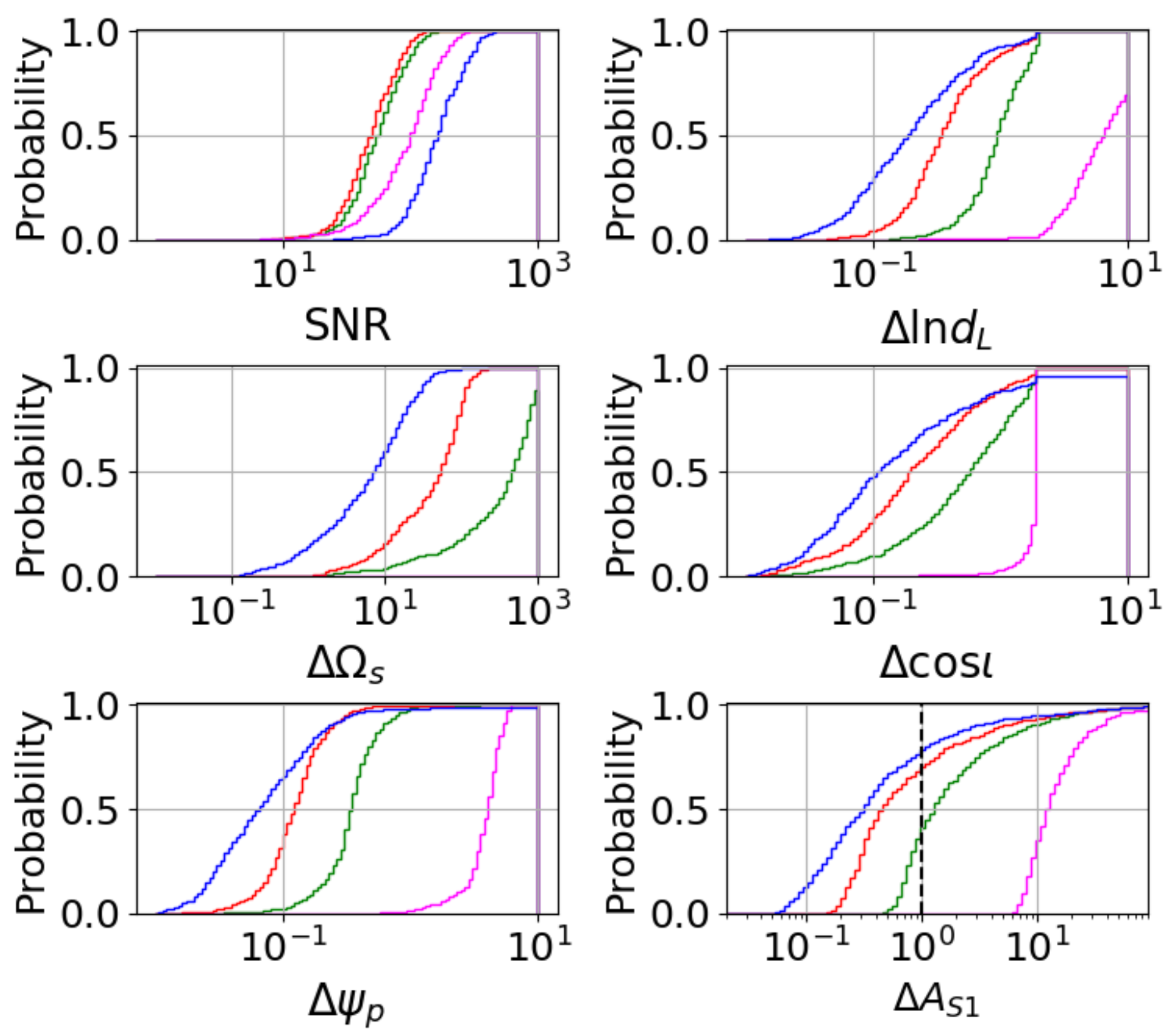}
 \end{center}
 \caption{Cumulative histograms for parameter estimation errors for $1.4M_{\odot}-1.4M_{\odot}$ BNS at $z=0.1$ in the model TS1 with a single 3G detector. The colors are with ET-B(green), with ET-D(red), CE(magenta), and with ideal(blue).}
 \label{ts1}
\end{figure}

The results of the parameter estimation for BNS with a single 3G detector are shown in \Table{table_result}.  We show the medians of parameter estimation errors of the luminosity distance, the sky localization, and the additional polarization amplitude. Polarization modes would be separable even with either an individual ET-D and an ideal detector, while they would be inseparable with either an ET-B and a CE. The sensitivity in the sub-5 Hz low-frequency range is crucial to localize the source position as reported \cite{Zhao2017, Chan2018}. The sensitivities of an ET-D and an ideal detector are better than those of an ET-B and a CE in the region. Thus, the time-varying effect of the antenna pattern functions helps us to better separate the polarization modes of GWs. 

\Fig{ts1} shows the cumulative histograms in the model TS1 for luminosity distance, the sky localization, the inclination angle, the polarization angle, and the additional polarization amplitude for all sources. It also indicates that the Earth's rotation can break the degeneracies among amplitude parameters. The errors of the additional polarization amplitude parameter for quite a few BNSs is less than unity with ET-B. This shows that the effect of the time-varying antenna pattern functions can help breaking the degeneracy among amplitude parameters partially. Thus, the polarizations would be separable for some BNS sources with ET-B depending on the positions of sources relative to the detector. A single ET-like 3G detector can be used to test polarizations by the observations of BNS.

We change the lower cutoff frequency $f_{\rm min}$ in the case of ET-D to search for the critical frequency region in the polarization test. The results are shown in \Table{table_result_fmin}. The frequencies of $1\unit{Hz}$, $5\unit{Hz}$, and $10\unit{Hz}$ correspond to $5.38\unit{days}$, $1.77 \unit{hours}$, and $0.28 \unit{hours}$  before the merger, respectively, for $1.4M_{\odot}-1.4M_{\odot}$ BNS from \Eq{tf}. The polarization modes would be inseparable with the lower cutoff frequency $f_{\rm min}=5\ {\rm Hz}$ even in the case of ET-D. This suggests that the sub-5 Hz range, especially the range below $5\ {\rm Hz}$ is essential for the polarization test. The 3G detectors are designed so that the lower frequency sensitivity is limited by the Newtonian noise, which is a noise caused by the fluctuations of the surrounding gravitational potential \cite{Saulson1984, Harms2015}. Therefore, it is important to gain better understanding about the Newtonian noise not only in localization of GWs but also in the test of GW polarizations with 3G detectors because the Newtonian noise is still poorly understood.

\begin{table*}
\caption{Medians of parameter estimation errors for $1.4M_{\odot}-1.4M_{\odot}$ BNS at $z=0.1$ when changing the lower cutoff frequency $f_{\rm min}$ in the case of ET-D. The range below 5 Hz is important for the polarization test.}
  \begin{tabular}{|c|c|c|c|c|} \hline
   & parameter & BNS(ET-D) fmin=1Hz & BNS(ET-D) fmin=5Hz & BNS(ET-D) fmin=10Hz \\ \hline
		 &SNR 							& 50.7 	& 49.8 		& 41.6 		\\  \hline 		  
		&$\Delta\ln{d_L}$ 					& 0.355	&  0.910 		& 2.31 		\\ 
ModelTS1&$\Delta\Omega_s[\rm{deg}^2]$ 		& 55.6 	&  184  		& 4308 		\\    
		&$\Delta A_{S1}$					& 0.459 	& 1.55 		& 4.56  		\\ \hline
  \end{tabular}
  \label{table_result_fmin}
\end{table*}

\begin{figure}
 \begin{center}
 \includegraphics[width=\hsize]{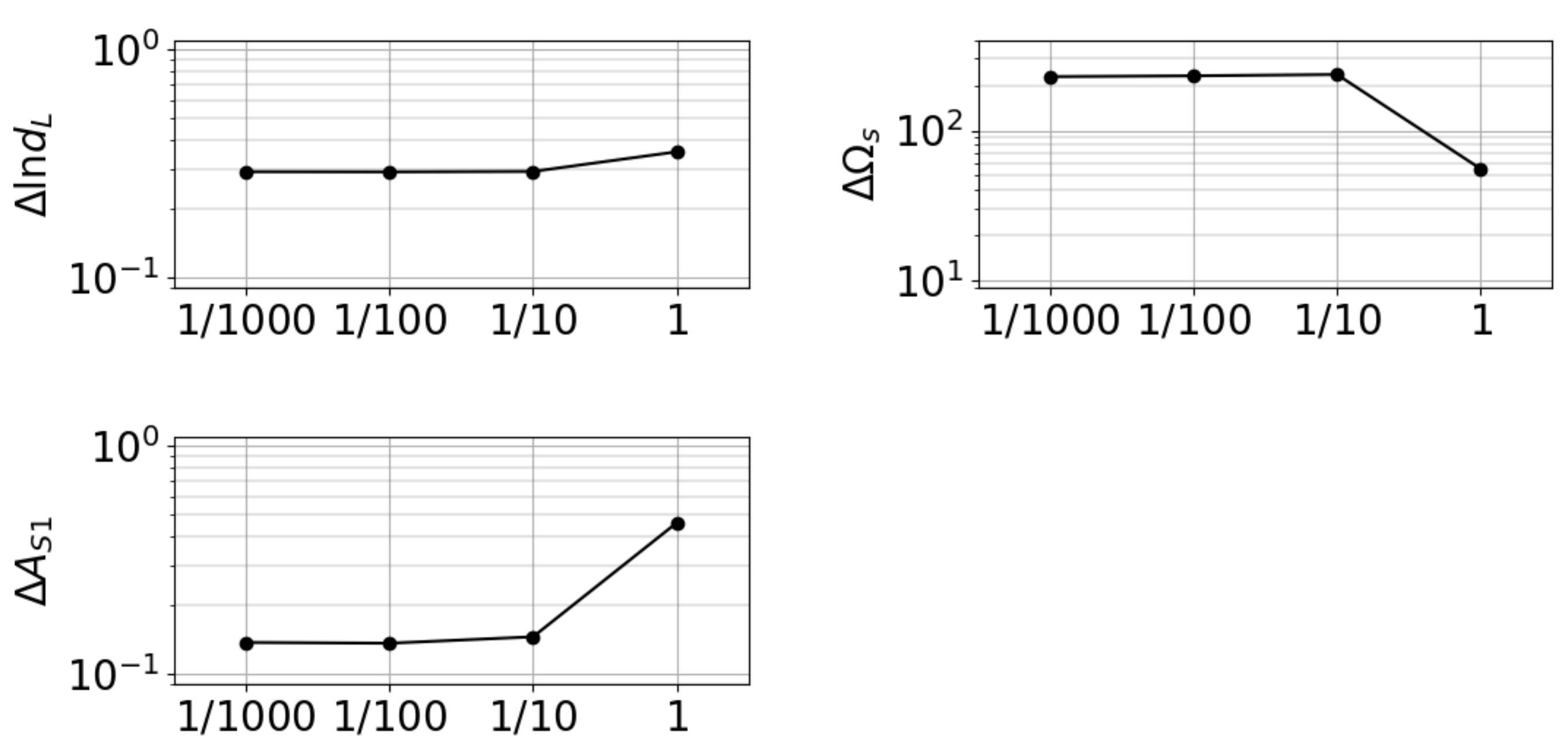}
 \end{center}
 \caption{Medians of parameter estimation errors for $1.4M_{\odot}-1.4M_{\odot}$ BNS at $z=0.1$ when changing the fiducial value of $A_{S1}$ in the case of an ET-D. The error of $A_{S1}$ is hardly  changed when the fiducial values are less than $1/10$, giving the detection limit.}
 \label{amp_diff_ts1_ET-D_01NS}
\end{figure}

For a proof-of-principle study, it is assumed that the fiducial value of the additional amplitude parameter for the scalar mode is unity in our analysis above to research the fundamental effect of the Earth's rotation in the polarization test. We studied how the choice of the fiducial values affect the estimation errors by changing the fiducial values to $1/1000, 1/100,$ and $1/10$. \Fig{amp_diff_ts1_ET-D_01NS} shows the fiducial-value dependence of the errors in the model TS1 with an ET-D. $A_{S_1}$ error is hardly  changed at the fiducial values less than $1/10$. The dependence of fiducial values is consistent with the results with a GW detector network formed by the 2G detectors \cite{Takeda2018}.  However, this detection limit is not given by $1/{\rm SNR}$, in contrast to the case of the 2G detectors. The reason would be a single 3G detector at each time forms an effective detector network, but does not form a real multi-detector network at the same time like the 2G detectors. 

In \Table{sens_10}, we checked whether a single 3G detector breaks the degeneracies among the amplitude parameters or not by artificially improving the detector sensitivity of an ET-D and a CE by a factor of $10$. As a result the error of the amplitude parameter is improved by $10^{0.998}$ with an ETD or is scaled by SNR, while the error is improved by only $10^{0.185}$ with a CE. This indicates that the effective network formed by a single ET-like detector could break the degeneracies among the amplitude parameters, but a CE-like detector could not take advantage of the effect of the Earth rotation enough due to its worse sensitivity at low frequencies. The index of the improvement in the error can be used to quantify how the degeneracies are broken, $0.998$ in the case of ET-D and $0.185$ in the case of CE. 

 \begin{table*}
\caption{Medians of parameter estimation errors for $1.4M_{\odot}-1.4M_{\odot}$ BNS at $z=0.1$ when changing the sensitivity better by a factor of 10. $A_{S1}=1$. The error of the additional amplitude parameter is improved by $10^{0.998}$ with ET-D and by $10^{0.185}$ with CE. }
  \begin{tabular}{|c|c|c|c|c|} \hline
   & parameter & BNS(ET-D) & BNS(CE) \\ \hline
		 &SNR 							& 507 	& 1047			\\  \hline 		  
		&$\Delta\ln{d_L}$ 					& 0.0357 	&  3.73 		 	\\ 
ModelTS1	&$\Delta\Omega_s[\rm{deg}^2]$ 		& 0.558 	&  12727  		 	\\    
		&$\Delta A_{S1}$					& 0.0461 	& 8.43 		 	\\ \hline
  \end{tabular}
  \label{sens_10}
\end{table*}

We also analyzed with three types of detector networks (ET-D - ET-D, ET-D - CE, CE-CE). Each two detectors is located at the Livingston site of aLIGO and Virgo site. Parameter estimation results are shown in  \Table{table_result_2}, taking the fiducial values of the additional polarization amplitudes unity. The ET-D - ET-D can separate the polarizations more accurately and the polarizations would be separable also with ET-D - CE. On the other hand, it would be possible to distinguish the polarizations even with two CE-like detectors to some extent, but its ability is significantly limited because CE detectors can hardly utilize the Earth rotational effect, although a CE detector can obtain high SNR due to its significantly better sensitivity above $10\unit{Hz}$ depending on the masses of compact binary.

\begin{table*}
\caption{Medians of parameter estimation errors for $1.4M_{\odot}-1.4M_{\odot}$ BNS at $z=0.1$ with three types of detector networks (ET-D - ET-D, ET-D - CE, CE-CE). Three polarization modes would be separable with networks composed of two 3G detectors.}
  \begin{tabular}{|c|c|c|c|c|c|} \hline
   & parameter & BNS(ET-D - ET-D) & BNS(ET-D - CE) & BNS(CE-CE) \\ \hline
		 &SNR 							& 75.2 	& 120 		& 151 	\\  \hline 		  
		&$\Delta\ln{d_L}$ 					& 0.0520 	&  0.124 		& 0.569 	\\ 
ModelTS1&$\Delta\Omega_s[\rm{deg}^2]$ 		& 0.346 	&  0.643 		& 3.51 	\\    
		&$\Delta A_{S1}$					& 0.0797 	& 0.178 		& 0.913 	\\ \hline
  \end{tabular}
  \label{table_result_2}
\end{table*}

\subsection{Binary black hole}
\label{BBH}

Next we consider polarization test with BBHs. We estimate model parameters for BBHs with equal $10M_{\odot}$ at z=0.5 in the model TS1 with a single 3G GW detector such as ET-B, ET-D, CE and an ideal detector. Angular parameters are uniformly random and 500 sources are calculated by the Fisher analysis as with the case of BNSs. \Table{table_result_3} shows the results of the parameter estimation for BBH with a single 3G detector. The medians of parameter estimation errors are shown. However, since the upper cutoff frequency for $10M_{\odot}-10M_{\odot}$ BBHs at z=0.5 is about $147\unit{Hz}$, the duration of the signal is shorter than that of BNS. Thus, it is relatively difficult for a single 3G detector to localize the BBH merger and determine the parameters in amplitude including the polarization parameter, that is, the polarizations would not be separable.

\begin{table*}
\caption{Medians of parameter estimation errors for $10M_{\odot}-10M_{\odot}$ BBH at $z=0.5$ with a single 3G GW detector. For BBH, it is relatively difficult for a single 3G detector to determine the parameters in amplitude including the polarization parameter due to the short duration of the signal. }
  \begin{tabular}{|c|c|c|c|c|c|} \hline
   & parameter & BBH(ET-B) & BBH(ET-D) & BBH(CE) & BBH(Ideal) \\ \hline
		 &SNR 							& 57.9 	& 51.6 		& 111 	& 181	\\  \hline 		  
		&$\Delta\ln{d_L}$ 					& 4.204 	&  3.390 		& 20.45 	& 2.181	\\ 
ModelTS1 &$\Delta\Omega_s[\rm{deg}^2]$ 		& 26437 	& 15618		& 219581	& 3133	\\    
		&$\Delta A_{S1}$					& 7.52 	& 6.36 		& 41.7 	& 4.34	\\  \hline
  \end{tabular}
  \label{table_result_3}
\end{table*}

\Table{table_result_4} shows the results for BBHs with two-detector networks (ET-D - ET-D, ET-D - CE, CE-CE). Three polarization modes from BBH mergers would be separable with two detectors including an ET-D. This indicates that the 3G detector networks can also take advantage of the Earth rotation effect for BBHs.

\begin{table*}
\caption{Medians of parameter estimation errors for $10M_{\odot}-10M_{\odot}$ BBH at $z=0.5$ with two-detector networks (ET-D - ET-D, ET-D - CE, CE-CE). Three polarization modes would be separable with networks composed of two 3G detectors including ET-D.}
  \begin{tabular}{|c|c|c|c|c|c|} \hline
   & parameter & BBH(ET-D - ET-D) & BBH(ET-D - CE) & BBH(CE-CE) \\ \hline
		 &SNR 							& 77.4 	& 126		& 161 	\\  \hline 		  
		&$\Delta\ln{d_L}$ 					& 0.062 	&  0.045 		& 3.12 	\\ 
ModelTS1&$\Delta\Omega_s[\rm{deg}^2]$ 		& 0.819 	&  2.79  		& 122 	\\    
 		&$\Delta A_{S1}$					& 0.089 	& 0.521 		& 6.45 	\\ \hline
  \end{tabular}
  \label{table_result_4}
\end{table*}

\subsection{Multiple sources}
\label{Multiple sources}

Finally,  we show the prospects for the GW polarization test by future observations of multiple compact binary mergers with the 3G ground-based detectors, considering their redshift distribution.

\subsubsection{Compact binary merger rate and detection rate}
\label{Prospects}
Let $R_m(z)$ be redshift rate density in the detector frame, that is the number of mergers per detector time per redshift,
\begin{equation}
R_m(z_m):=\frac{dN_m}{dt dz}=\frac{1}{1+z_m}\frac{dV_c}{dz}\mathcal{R}_m(z_m),
\label{redshift_rate_density_in_the_detector_frame1}
\end{equation}
where $N_m$ is the number of mergers, $t$ is the detector frame time, $V_c$ is the comoving volume, and $z_m=z(t_m)$ is the redshift of the binary system that merges at a lookback time $t_m$ \cite{Vitale2018}. Here $\mathcal{R}_m$ is the volumetric merger rate in the source frame, that is the number of the mergers per comoving volume per source time, defined by
\begin{equation}
\mathcal{R}_m(z_m):=\frac{dN_m}{dV_c dt_s},\\
\label{volumetric_merger_rate_in_the_source_frame1}
\end{equation}
where $t_s$ is the source frame time, 

The volumetric merger rate depends on the binary formation rate and the delay time distribution between the formation of the binary system and their coalescence. 
\begin{equation}
\mathcal{R}_m(t_m)=\int^{\infty}_{t_m}dt_f\mathcal{R}_f(t_f)p(t_m| t_f; \lambda_f),
\end{equation}
where $\mathcal{R}_f$ is the binary formation rate and the delay time distribution , $p(t_m | t_f; \lambda_f)$ is the probability density that a binary system formed at time $t_f$ merge at time $t_m$. The delay time distribution may depend on some parameters $\lambda_f$, for example some kind of time scale parameters and the parameters of the merging binary system.
This also can be expressed in terms of redshift,
\begin{equation}
\mathcal{R}_m(t_m(z_m))=\int^{\infty}_{z_m}dz_f\frac{dt_f}{dz_f}\mathcal{R}_f(t_f(z_f))p(t_m| t_f; \lambda_f),
\label{volumetric_merger_rate_in_the_source_frame2}
\end{equation}
where $z_f=z(t_f)$ is the redshift of the binary system that forms at a lookback time $t_f$.

It is assumed that the volumetric binary formation rate $\mathcal{R}_f(z_f)$ is simply proportional to the star formation rate density at the same redshift, Madau plot $\psi(z)$ \cite{Madau2014}. The formation rate or the delay time distribution may depend on some properties of the formed binaries. However, it is also assumed here that the dependence of the intrinsic parameters can be ignored for simplicity. We consider the distribution of delay time between formation and merger uniform in the logarithm of the time delay,

\begin{equation}
\begin{split}
p(t_m | t_f)&=p(\log(t_m-t_f))\\ &\propto \begin{cases}
    1 & (10 \unit{Myr}<t_f-t_m<10 \unit{Gyr}) \\
    0 & (10 \unit{Myr}>t_f-t_m \lor t_f-t_m>10 \unit{Gyr}).
  \end{cases}
  \end{split}
\end{equation}

\subsubsection{Prospects for polarization test with future ground-based detectors}
We produce the source catalog that reflects the above binary merger rate distribution. The following is our procedure to create the catalog. First we calculate the volumetric merger rate in the source frame $\mathcal{R}_m(z_m)$ using \Eq{volumetric_merger_rate_in_the_source_frame2}, then calculated the redshift rate density in the detector frame $R_m(z)$ using \Eq{redshift_rate_density_in_the_detector_frame1}. We get the binary merger rate at redshift $z$ with redshift-bin $z=0.1$ by integrating \Eq{redshift_rate_density_in_the_detector_frame1}. We produced each $5000$ sources of  $1.4M_{\odot}-1.4M_{\odot}$ BNS and $10M_{\odot}-10M_{\odot}$ BBH having the above redshift merger distribution and estimated their parameters for the inspiral waveforms in the model TS1. The fiducial values of the additional polarization amplitude parameters are set to be enough small, $A_{S1}=1/1000$. The median values of the errors of the additional polarization amplitude parameter at each redshift $\Delta A_{S1} (z)$ are obtained by evaluating the middle number when the errors in the same redshift bin are arranged from lowest to highest. It is expected that the actual errors could be improved statistically from future observations of  binary mergers. We could assume that they improve inversely proportional to square root of the number of detections $N_z$ at the redshift $z$ statistically,
\begin{equation}
\Delta A_{s,S1}(z)=\frac{\Delta A_{S1} (z)}{\sqrt{N_z}}.
\end{equation}
We calculate the expected number of detections $N_z$ in a redshift bin from $z-\Delta z$ to $z$ (we take $\Delta z=0.1$) as the product of the number of mergers $N_{m,z}$ in the same redshift bin and detection probabilities at each redshift $p_d(z)$,
\begin{equation}
N_z=N_{m,z}p_d(z).
\end{equation}

The detection probability $p_d(z)$ is the ratio of the number of detections to the number of binary mergers at a redshift that are estimated from SNR calculations in which the detection criterion is ${\rm network\ SNR} > 8$. \Table{statistical_procedure} summarizes these procedures in the case of BNS.

\begin{table*}
\caption{Parameter estimation results for $1.4M_{\odot}-1.4M_{\odot}$ BNS with an ET-D in a redshift bin from $z-\Delta z$ to $z$ ($\Delta z=0.1$). Observation period is chosen such that a neutron binary star merger occurs within the distance of $z = 0.01$ during the observation, corresponding to $1.82\ {\rm yr}$ when assuming the binary merger rate for binary neutron star $1540\ {\rm Gpc^{-3}yr^{-1}}$.}
  \begin{tabular}{|c|c|c|c|c|c|c|c|c|c|c|} \hline
   & z=0.1 & z=0.2 & z=0.3 & z=0.4 & z=0.5 & z=0.6 & z=0.7 & z=0.8 & z=0.9 & z=1.0 \\ \hline
   Detection Probability $p_d$ &1&0.97&0.92&0.85&0.77&0.58&0.48&0.38&0.31&0.28\\ \hline   
   Merger rate $N_{m,z}$ &120   &525 &1197  &2088  &3176  &4422  &5779  &7198  &8627  &10013 \\ \hline
   Detection rate $N_z$ &120 &508&1102&1769&2460&2574&2795&2762&2709&2788 \\ \hline
   $\Delta A_{S1}/\sqrt{N_z}$ [$\times 10^{-2}$] &1.22 &1.16&1.14&1.18&1.23&1.49&1.63&1.95&2.11&2.43\\ \hline
  \end{tabular}
  \label{statistical_procedure}
\end{table*}

\begin{figure}
 \begin{center}
 \includegraphics[width=\hsize]{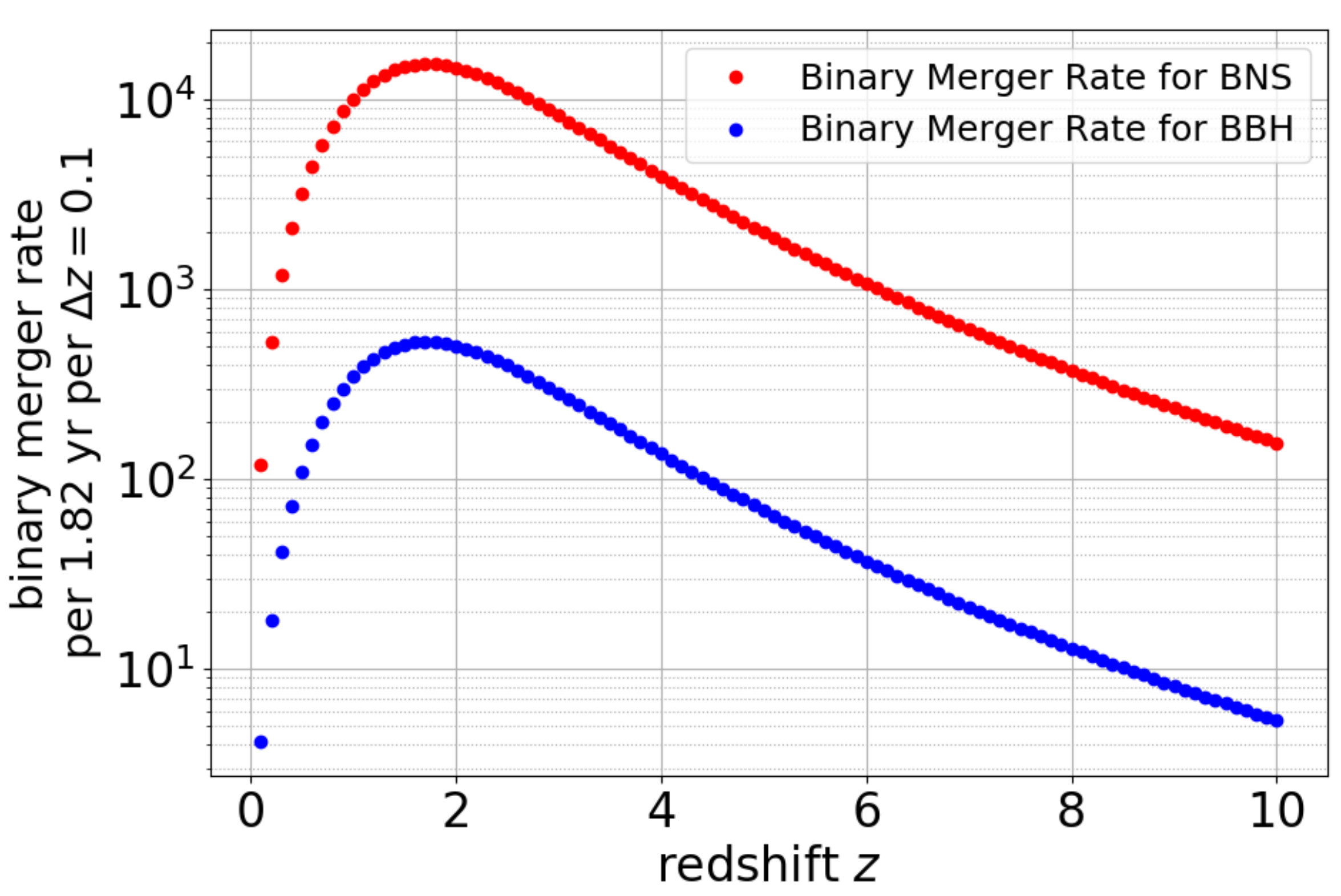}
 \end{center}
 \caption{Compact binary merger rate distribution based on Madau plot with the flat-in-log delay-time distribution. The overall factors of the star formation rate density are based on the BNS merger rate $1540\ {\rm Gpc^{-3}yr^{-1}}$ \cite{Abbott2017} and the BBH merger rate $53.2\ {\rm Gpc^{-3}yr^{-1}}$ \cite{TheLIGOScientificCollaboration2018}.}
 \label{flat_log}
\end{figure}

 Here the expected number of mergers $N_{m,z}$ for BNS is evaluated by the constructed distribution above where the overall factor of the star formation rate density is chosen such that a BNS merger occurs within the distance of $z=0.01$ during the observation. This corresponds to the observation period of 1.82 yr when the BNS merger rate is taken as $1540\ {\rm Gpc^{-3}yr^{-1}}$ \cite{Abbott2017}. The overall factor of the star formation rate density for BBH is determined by assuming that  the binary merger rate for BBH is $53.2\ {\rm Gpc^{-3}yr^{-1}}$ \cite{TheLIGOScientificCollaboration2018}. \Fig{flat_log} shows our binary merger distribution.

Finally, we find the expected error of the additional polarization amplitude parameter from future observations from the equation,
\begin{equation}
\frac{1}{\Delta A_{F, S1}^ 2}=\sum_{z} \frac{1}{\Delta A_{s, S1}(z)^2},
\end{equation}
because we consider independent events.

The result of the expected error from future observations of $1.4M_{\odot}-1.4M_{\odot}$ BNS with an ET-D is
\begin{equation}
\Delta A_{F, S1, ET-D}=4.45\times10^{-3}.
\end{equation}
On the other hand, the result of the expected error from future observations of $10M_{\odot}-10M_{\odot}$ BBH with ET-D - CE is 

\begin{equation}
\Delta A_{F, S1, ET-D - CE}=1.21\times10^{-3},
\label{ETDCE}
\end{equation}
and that with ET-D - ET-D is 
\begin{equation}
\Delta A_{F, S1, ET-D - ET-D}=4.66\times10^{-4}.
\label{ETDETD}
\end{equation}

\section{Discussions}
\label{Discussions}
\subsection{Nearby event vs multiple sources}
In the previous section, we considered the improvement of the measurement errors by combining multiple sources. However, they should include very nearby events, which may contribute dominantly to the determination of the polarization amplitude parameter. To clarify at which redshift the polarization amplitude parameter is determined well, we also calculated the medians of parameter estimation errors of the additional polarization amplitude parameter for a single source of BNS at $z = 0.01$ to compare with  above expected values. The median of the error is
\begin{equation}
\Delta A_{S1}=1.40\times10^{-2}.
\end{equation}
This suggests that an event near the Earth ($z\simeq0.01$) is available to test the polarizations with a precision of $\Delta A_{S1}\simeq1.40\times10^{-2}$, but it would be possible to test by the observations of multiple binary systems with precisions of $4.45\times10^{-3}$ or $1.21\times10^{-3}$ within the period during which such a nearby-BNS merger event would occur at least once. For one nearby event, the error gets worse with the luminosity distance as $\Delta A_{S1}\sim {\rm 1/SNR} \propto d_L$, while for multiple sources in a redshift bin the error is estimated as $\Delta A_{S1}\sim {\rm 1/SNR}/\sqrt{N}\propto d_L/\sqrt{d_{L}^2} \sim {\rm constant}$. Thus, it is expected that the total expected estimation error $\Delta A_{S1}$ is reduced by the square root of the number of the bin combining multiple sources.

We assume the statistically improvement of the error $\Delta A_{S1}$ for multiple sources in the same redshift bin. A specific method to deal with observations of the compact binary coalescences statistically need to be developed in the future. For example,  a stacking method could be applied to look for a small signal of the nontensorial polarization modes \cite{Yang2017}.  

\subsection{Limitation by PSR B1913+16 }
The amplitude for the nontensorial mode has already limited by the observation of PSR B1913+16. The orbital energy $E_{\rm orbit}$ of a binary system is related to the frequency of a radiated GW \cite{Maggiore2007} by
\begin{equation}
E_{\rm orbit}=-\left(\frac{\pi^2\mathcal{M}^5f^2}{8}\right)^{1/3},
\end{equation}
at Newtonian order.
Then,
\begin{equation}
\frac{\Delta \dot{E}_{\rm orbit}}{\dot{E}_{\rm orbit}}=\frac{\Delta \dot{f}}{\dot{f}}=\frac{\Delta \dot{P}}{\dot{P}},
\end{equation}
where $P$ is the orbital period of a binary system and the dot notation for time differentiation is used. The observation of the orbital period of PSR B1913+16 has constrained $\dot{P}$ \cite{Will2005},
\begin{equation}
\frac{\Delta \dot{P}}{\dot{P}}=0.003 \pm 0.002.
\end{equation}
From $\dot{E}_{\rm orbit}\propto A^2$, a possible modification of the GW amplitude for the nontensorial mode is limited by
\begin{equation}
A_{S1} \lesssim 1\times10^{-3}.
\end{equation}
Thus, the observations of compact binary coalescences by the 3G detectors would be able to reach the compatible level to the current constraint by a more direct and robust way. It is noted that the current limitation from PSR B1913+16 has been imposed much before a compact binary merger, that is, in a much weaker gravity regime.


\subsection{Distributions for multiple sources}
For simplicity, we ignored some distributions of BHs or NSs, for example, mass, mass ratio and so on when we consider multiple sources. A CE-like detector could detect CBC at higher redshifts due to its significantly better sensitivity depending on the masses of the source. However, not only the separability of polarizations of an ET-D is better than that of a CE due to its better lower sensitivity, but also the inspiral range of ET-D is higher than that of CE for $10M_{\odot}-10M_{\odot}$ BBH \cite{Hall2019}. Thus, the error with ET-D - ET-D network \Eq{ETDETD} is less than that with ET-D - CE network \Eq{ETDCE} for multiple BBH sources. 

For a proof-of-principle study here, we did not consider the formation channels other than isolated field binaries. BBH formed in globular clusters may contribute to the compact binary formation rate \cite{Rodriguez2015, Fujii2017, Fragione2018, Rodriguez2018}. However, since the redshift dependence of the formation rate of BBH formed in globular clusters would have similar dependence of the star formation rate and it is assumed that the observational merger rate by LIGO and Virgo includes the contributions from BBH formed in globular clusters, our analysis would also include the contributions from BBH formed in globular clusters.
On the other hand,  the binary stars from population III stars may also contribute to the compact binary formation rate \cite{Kinugawa2014, Belczynski2016}. Population III stars are not likely to become BNS, but they form BBH at high redshift $z>6$. If we include contributions from such a specific formation channel having different redshift distribution from the star formation rate, the error may be reduced when the detection number would increase. The contributions from population III stars would be important in the case of BBH because the 3G detectors such as ET-D and CE have large inspiral range.


\subsection{Experimental aspects}
The frequency range below 5 Hz is essential in the polarization test as we mentioned. From the point of view of a detector, the realization of the sensitivity at lower frequencies is challenging. One of the dominant noise sources at the lower frequency region below $30\unit{Hz}$ is fluctuations in a local gravity field caused by moving objects around the test masses of a detector, which is called Newtonian noise \cite{Saulson1984, Harms2015}. Seismic Newtonian noise \cite{Hughes1998} and atmospheric Newtonian noise \cite{Fiorucci2018} can be dominant in the future detector. For seismic Newtonian noise, some cancellation schemes  by an array of seismometers or tiltmeters \cite{Driggers2012, Coughlin2016, Harms2016} and passive suppression method \cite{Harms2014} have been proposed. On the other hand, local gravity-noise suppression by constructing the detector underground is proposed for atmospheric Newtonian noise \cite{Fiorucci2018}. 

\section{Conclusion}
\label{Conclusion}
We studied separability and the effect of the Earth rotation in the polarization test with compact binary coalescences by the 3G detectors such as ET and CE via a parameter estimation approach.   \Table{result_summary} shows summary of our results.

\begin{table*}
\caption{Summary of the separability of polarization modes of GWs from compact binary coalescences with 3G detectors for $1.4M_{\odot}-1.4M_{\odot}$ BNS at $z=0.1$ or $10M_{\odot}-10M_{\odot}$ BBH at $z=0.5$ in the model TS1 where there are two tensor modes and a scalar mode. The median values of the additional polarization amplitude parameter $A_{S1}$ are shown when the fiducial values are unity $A_S=1$. }
  \begin{tabular}{|c|c|c|c|c|c|c|c|} \hline
   & ET-B & ET-D & CE & ETD -ET-D & ETD - CE & CE - CE\\ \hline
	BNS			&inseparable	&separable 	&inseparable	&separable	&separable	&separable  	\\ 
	$\Delta A_{S1}$	&			&0.459		&			&0.0797		&0.178		&0.913 \\ \hline 
	BBH			&inseparable	&inseparable	&inseparable	&separable 	&separable	&inseparable	\\  
	$\Delta A_{S1}$	&			&			&			&0.089		&0.521		&  \\ \hline 
  \end{tabular}
  \label{result_summary}
\end{table*}

In the case of BNSs, an ET-D could separate the polarizations even with a single detector because better sensitivity at frequencies lower than $5\unit{Hz}$ can take advantage of the Earth's rotation. Time-varying antenna pattern functions help with forming an effective detector network along the trajectory of a detector. This effective network makes it possible to test polarizations with a single 3G detector. On the other hand, it would be difficult for a single CE-like detector to separate polarizations because the low frequency sensitivity is not better than that of an ET-D. In the case of BBHs, a single 3G detector could not distinguish three polarizations due to the low upper cutoff frequency for BBHs resulting in the short duration of the signal. However, 3G detectors can make use of a part of time-varying effect of the antenna patterns. Thus, three polarizations from BBH merger would be separable with a detector network composed of  two 3G detectors including an ET-D.

We also studied the prospects for the GW polarization test with multiple sources based on the compact binary merger distributions. A single golden event, BNS merger at $z\simeq0.01$, could be used to test polarizations with precisions of  $1.40\times10^{-2}$.  On the other hand, it is expected that it would be possible to test the polarizations by the observations of binary systems with accuracy $4.45\times10^{-3}$ for BNSs or $1.21\times10^{-3}$ for BBHs within the observational period during which such a golden event would occur. These precisions are comparable to the current constraints on the amplitude for nontensorial modes from the observations of PSR B1913+16, though the current constraint has been imposed at the stage much before a binary merger, that is, in a much weaker gravity regime. 

\section*{Acknowledgements}
The authors would like to thank T. Kinugawa and T. Shimoda for frutiful discussions. H. T. and K. K. acknowledge financial support received from the Advanced Leading Graduate Course for Photon Science (ALPS) program at the University of Tokyo. H.T., A.N., K.N., K.K. and Y.M.  are supported by JSPS KAKENHI Grant No. 18J21016., JSPS KAKENHI Grant No. JP17H06358. and No. 19H01894., JSPS KAKENHI Grant  No. 17J01176,JSPS KAKENHI Grant No. 16J01010 and JSPS Grant-in-Aid for Scientific Research (B) No. 18H01224, respectively. \newpage


\bibliography{paper_pol_3G}
\bibliographystyle{h-physrev3}
\end{document}